\documentstyle[12pt,epsf]{article}
\def\D{\partial}
\def\DF#1#2{\frac{\partial #1}{\partial #2}}
\def\DDF#1#2{\frac{\partial^2 #1}{{\partial#2}^2}}
\def\N{\nabla}
\def\NN{\nabla^2}
\def\RS{{\rm S}}
\def\NS{\nabla\!_{\RS}}
\def\NNS{{\NS\!}^2}
\def\intS{\oint_\RS}
\def\mib#1{\mbox{\boldmath$#1$\unboldmath}}
\def\be{\begin{equation}}
\def\ee{\end{equation}}
\def\ba{\begin{eqnarray}}
\def\ea{\end{eqnarray}}
\def\nn{\nonumber}
\def\tu{\tilde{u}}
\def\sigind{\sigma_{\rm ind}}
\begin{document}
\title{{\Large An Explicit Form of the Equation of Motion of the Interface in 
Bicontinuous Phases}}
\author{Hiroyuki TOMITA\\
        Department of Fundamental Sciences, \\
        Faculty of Integrated Human Studies,\\
        Kyoto University, Kyoto 606-8501, Japan\\
        (Email: tomita@phys.h.kyoto-u.ac.jp)}

\maketitle
\begin{abstract}
  The explicit form of the interface equation of motion derived assuming
a minimal surface is extended to general bicontinuous interfaces that appear
in the diffusion limited stage of the phase separation process of binary
mixtures.
  The derivation is based on a formal solution of the equivalent simple layer
for the Dirichlet problem of the Laplace equation with an arbitrary boundary
surface.
  It is shown that the assumption of a minimal surface used in the previous
linear theory is not necessary, but its bicontinuous nature is the essential
condition required for us to rederive the explicit form of the simple layer.
  The derived curvature flow equation has a phenomenological cut-off length,
i.e., an `electro-static' screening length.
  That is related to the well-known scaling length characterizing
the spatial pattern size of a homogeneously growing bicontinuous phase.
  The corresponding equation of the level function in this scheme is given
in a one-parameter form also.
\end{abstract}

\section{Introduction}
   The ordering process associated with the first order phase transition
has been investigated for a long time since the early works by van der 
Waals appeared at the end of the last century.\cite{VDW}\ 
   It has occupied a major position together with the dynamical critical
phenomena as a subject not only of the non-equilibrium statistical physics
but also of the field theory.
   A challenging feature of it is the strong nonlinearity which causes a
variety of evolving spatial patterns.
   This has initiated the interesting concept of `pattern formation' in 
far-from-equilibrium thermodynamic systems.\cite{GD}-\cite{BRAY}

\par
   Complicated patterns, at a glance, are seen even in the most simplified
theoretical models of a binary mixture, e.g., in the time-dependent
Ginzburg-Landau (TDGL) model for the conserved order parameter (COP),
or the Cahn-Hilliard (CH) model.\cite{CH}\ 
   The nonlinearity in these systems finally gives rise to a spatial
singularity of the order parameter gap, i.e., the interface with a sharp
profile that separates distinctly the coexisting couple of phases.
   This is a primary example of the so-called `topological defect'
appearing in continuous fields.
   The main subject for us in this stage is to reduce a degenerate evolution
equation for the interface from the basic TDGL or CH equations for the bulk
field.

\par
   In spite of remarkable success of the intuitive, droplet theories on
a dilute mixture, i.e., some pioneering works in the middle of this century,
\cite{TODES}-\cite{WAGNER}\ the interface evolution equation for more
general systems has not been obtained yet in an explicit form.
   The most general form we have found so far is an integral equation
obtained by Kawasaki and Ohta.\cite{KO,KOP}\ 
   This is a kind of the curvature flow equation which represents the
interface velocity with the mean curvature {\it implicitly} in contrast
to the Allen-Cahn equation\cite{AC} for a system of non-conserved order
parameter (NCOP).
   It has been shown\cite{TPR} that this is the equivalent simple layer
equation\cite{CHIL} for the Dirichlet problem of the Laplace equation that is
derived as a quasi-static approximation for a Stefan problem,\cite{STEFAN}
\ i.e., a diffusion problem of an autonomous boundary condition evolving
through the diffusion process itself.
   For some simplified systems, e.g., a planar interface or a thin system
of spherical droplets, it is straightforward to derive the curvature flow
equations, if they are considered as electro-static potential problems.
   The explicit potential solutions for such systems can be found
in every elementary textbook on electro-statics.\cite{JACKSON}\ 
   On the contrary, we will soon come up against a serious difficulty of
a complicated geometry of the boundary condition when we try to find out
explicit potential solutions for more general systems.
   For example, in a nearly symmetric mixture there appears a random
bicontinuous structure, i.e., the so-called {\it sponge} phase.
   This has an infinitely multiply-connected topology and is characteristic
of a three dimensional system.
   That contrasts to the case of a lamellar phase, which is essentially
equivalent to a planar system.

\par
   In the previous work\cite{TPR} the author assumed an almost minimal
surface for such a system mainly by intuition gained from some simulated
pictures\cite{CG}-\cite{KOGA}, which just remind us a periodic minimal
surface such as the Schwarz lattice structure,\cite{MINIMAL}\ at least
in a local view.
   The theoretical foundation for this assumption was that the mean
curvature is included linearly in the boundary value itself as the
Gibbs-Thomson condition already, and then it can be neglected within
the limits of a linear theory with respect to the small mean curvature.
   This situation is similar to that of the decay process of
a perturbative deformation on a planar interface.
   We used the common feature of zero mean curvature of the unperturbed
basis in both systems.
   An important difference is that the sponge phase has an evident
characteristic length, which was used effectively as the mean
{\it electro-static} screening length in the previous work.

\par
   In the present paper it is shown that the assumption of a minimal
surface on the unperturbed basis is not necessary in deriving an explicit
solution from the integral equation and the same one-parameter curvature flow
equation is rederived for more general bicontinuous systems.
   We need the bicontinuous nature of the interface only.
   Here `bicontinuous' means that the interface divides the whole
$\mib{R}^3$ space into a couple of ({\it not} necessarily symmetric)
subspaces connected in each like a planar interface does, and not
into subspaces more than three.
   This is the only condition required to make up a formal expression for
the equivalent simple layer in terms of the potential problem.

\par
   In \S 2 the integral equation for the interface velocity is rederived
on the basis of Onsager's variational principle.
   It is shown that the principle of minimum dissipation is an example
of the so-called gradient dynamics and is very useful for such an
actual purpose to reduce the degenerate interface equation of motion
correctly from the basic transport equation for the bulk field.
   In \S 3 a formal explicit expression for the simple layer on an arbitrary
connected surface is derived referring to a solvable problem of a plane
boundary.
   The equation of the level function that corresponds to the present
curvature flow equation is discussed in \S 4.
\section{The Onsager Principle and Basic Equations}
   In the original work by Kawasaki and Ohta\cite{KO,KOP} a kind of the 
path-integral method and a variational principle associated with it were used.
   It seems instructive for us to follow it in terms of the familiar
Onsager principle of minimum dissipation\cite{ONS} with reference to the recent
gradient dynamics.
   To begin with, let us survey the Onsager variational principle.
\par
   Let $\mib{x}=\{x_i\}$ be a set of generalized thermodynamic variables,
their phenomenological transport equations being given by
\be \label{PHENOMENO}
\dot{x}_i = \sum_{j} L_{ij}X_j ,
\ee
where $\{L_{ij}\}$ are the Onsager coefficients and $\mib{X}=\{X_{i}\}$ is
a set of generalized thermodynamic forces, which are defined
by using the {\it entropy} function $S$ as
\be
\mib{X}=\N_{\!\mib{x}}S~~{\rm or}~~X_i = \DF{S}{x_i} .
\ee
   Onsager proposed a minimum principle on the basis of the symmetric
property of $\{L_{ij}\}$, i.e., $L_{ij}=L_{ji}$, and its positive
definiteness as follows:
   Define two kinds of dissipation function and a Lagrangian by
\be
\Phi[\dot{\mib{x}},\dot{\mib{x}}]=\frac{1}{2}\sum_{i,j} {L^{-1}}_{ij}
\dot{x}_i\dot{x}_j ,
\ee

\be
\Psi[\mib{X},\mib{X}]=\frac{1}{2}\sum_{i,j}L_{ij}X_i X_j,
\ee
and
\ba
{\cal L}[\dot{\mib{x}},\mib{X}]
&=& \Phi[\dot{\mib{x}},\dot{\mib{x}}]
                                -\dot{S}[\dot{\mib{x}},\mib{X}]
                                +\Psi[\mib{X},\mib{X}]\nn\\
&=& \frac{1}{2}\sum_{i,j}{L^{-1}}_{ij}(\dot{x}_i-\sum_k L_{ik}X_k)
(\dot{x}_j-\sum_{l}L_{jl}X_l),
\ea
where
\be
\dot{S}[\dot{\mib{x}},\mib{X}]=\sum_i \dot{x}_iX_i.
\ee
Then the phenomenological equation Eq.(\ref{PHENOMENO}) is given by
\be
\delta{\cal L}=0~({\rm minimum})~~{\rm with~respect~to~~} \dot{\mib{x}}.
\ee
   This variational principle was confirmed later as a problem of the most
probable path in the path integral method of the linear fluctuation
theory.\cite{OM,HASHI}\ 
   It has been understood widely that the principle itself is nothing but
a formal theory as is suggested in the above survey and is no use for an
actual purpose to find out the transport equation itself for a given system.
\par
   Recently, mathematicians have introduced a notion of `gradient dynamics'.
\cite{CT}\ 
   In this sense, the above Onsager's phenomenology is a Lagrange multiplier
version\footnote{The original plan is that a dynamics is to be defined
by maximizing $\dot{S}=\dot{x}\cdot\!\N_{\!\mib{x}} S$ with a constraint of
a properly chosen inner product $(\dot{\mib{x}},\dot{\mib{x}})=$~constant.}
of a gradient dynamics defined by an inner product,
\be
(\mib{u},\mib{v})=\sum_{i,j}{L^{-1}}_{ij} u_i v_j 
=2\Phi[\mib{u},\mib{v}].
\ee
   Mathematicians may define an arbitrary dynamics by using an arbitrary
inner product.
   That gives us a good hint:
   That is, it can be reasonably expected that we may find out the correct
interface dynamics easily, when we rewrite the inner product, i.e., the
physical dissipation function for the bulk into an interfacial form.

\par
   Let $\{s(\mib{r})\}$ be the scalar field whose evolution is described by
the TDGL equation of COP type, or equivalently the CH equation,
\be \label{TDGL}
\DF{}{t} s(\mib{r},t)=-L\NN\frac{\delta}{\delta s(\mib{r})}~S(\{s(\mib{r})\}).
\ee
In this case $S(\{s(\mib{r})\})$ is not the entropy but related to the free
energy functional $F(\{ s(\mib{r})\})$ by $S=-F/kT$, where $T$ is the
temperature and $k$ the Boltzmann constant and
\be
F(\{ s(\mib{r})\}) = \int d\mib{r} \frac{1}{2\chi_0} 
\Bigl\{ -\frac{1}{2} s(\mib{r})^2 + \frac{1}{4{s_0}^2} s(\mib{r})^4
+\xi^2 [\N s(\mib{r}) ]^2 \Bigr\} .
\ee
The equation (\ref{TDGL}) has a uniform equilibrium solution,
\be
s(\mib{r})=\pm s_0 ,
\ee
and a planar interface one, i.e., the so-called kink solution,
\be \label{KINK}
s_{\rm K}(z) =s_0\tanh\frac{z}{2\xi},
\ee
where $z$ is a normal coordinate perpendicular to the interface.
  Here $\xi$ corresponds to the thickness of the interface.
  The parameter $\chi_0$ is related to the susceptibility by
\be
{\chi_0}^{-1} = \frac{\delta^2 F}{\delta s(\mib{r})^2}\Big|_{\pm s_0}.
\ee
   From Eq.(\ref{KINK}) we get an expression of the surface tension,\cite{VDW}
\be \label{TENSION}
\sigma = \frac{\xi^2}{\chi_0}\int_{-\infty}^{\infty}(\N s_{\rm K})^2dz
=\frac{\xi(2s_0)^2}{6\chi_0},
\ee
as the excess free energy stored in the interface layer where $s\ne \pm s_0$.

\par
   Now let us derive the interfacial version from the bulk dissipation
function defined by
\be \label{PHI}
\Phi=\frac{1}{2L}\int\!\!\int G_0(\mib{r}-\mib{r}')
\DF{s(\mib{r})}{t} \DF{s(\mib{r}')}{t} d\mib{r} d\mib{r}',
\ee
where $G_0(\mib{r}-\mib{r}')$ is the inverse of the Onsager coefficient
$-\NN$ in Eq.(\ref{TDGL}) defined by
\be
\NN G_0(\mib{r}-\mib{r}')=-\delta(\mib{r}-\mib{r}') ,
\ee
which is the Green function, i.e., the Newton (or the Coulomb) potential.
   If we assume that the every element $dA$ of the interface S at $\mib{a}$
propagates without deforming its profile, it is straightforward to show that
\be
\int_{-\infty}^{\infty}\DF{s(\mib{r})}{t} dz = -2s_0 v_n(\mib{a}).
\ee
from a geometrical consideration, where $v_n(\mib{a})$ is the interface
normal velocity.
   Thus, the interface version of the dissipation function Eq.(\ref{PHI})
is given by
\be
\Phi=\frac{(2{s_0})^2}{2L}\intS\!\intS G_0(\mib{a}-\mib{a}')
v_n(\mib{a})v_n(\mib{a}')dAdA' ,
\ee
where the volume elements $d\mib{r}$ and $d\mib{r}'$ in Eq.(\ref{PHI}) are
replaced by $dzdA$ and $dz'dA'$.
   On the other hand, by definition of the surface free energy $\sigma$,
the interface version of $\dot{S}$ is written as
\be
\dot{S}=-\frac{1}{kT}\intS\sigma H(\mib{a})v_n(\mib{a}) dA ,
\ee
according to the formula on the change of the surface area caused by the
normal displacement $\delta z$ of the surface S,
\be
\delta \intS dA = \intS H(\mib{a}) \delta z dA ,
\ee
where $H(\mib{a})$ is the mean curvature of the surface at $\mib{a}$.
   Then, by using the variational principle $\delta(\Phi-\dot{S})=0$ with
respect to $v_n(\mib{a})$ we obtain an integral equation,
\be \label{INTEGRAL}
\intS G_0(\mib{a}-\mib{a}')v_n(\mib{a}')dA' =-\frac{D\xi}{6}~
[H(\mib{a})-\overline{H}] ,
\ee
which is just the equation obtained by Kawasaki and Ohta.
   Here a new notation,
\be
   D = \frac{L}{kT\chi_0}
\ee
is introduced for the simplicity together with the relation Eq.(\ref{TENSION}).
   Note that a constant, $\overline{H}$, is incorporated as the Lagrange
unknown multiplier in the present variational principle, corresponding to
the constraint of order parameter conservation,
\be \label{CONSERVATION}
\intS v_n(\mib{a})dA=0 .
\ee
   If the inverse $\Gamma_0(\mib{a},\mib{a}')$ of the integral kernel
$G_0(\mib{a}-\mib{a}')$ satisfying
\be
\intS \Gamma_0(\mib{a},\mib{a}'')G_0(\mib{a}''-\mib{a}')dA''=\delta(\mib{a}-\mib{a}'),
\ee
were given, the constant $\overline{H}$ would be determined by a weighted
average,
\be \label{HBAR}
\overline{H}=\intS\!\intS \Gamma_0(\mib{a},\mib{a}')H(\mib{a}')dAdA'
/\intS\!\intS \Gamma_0(\mib{a},\mib{a}')dAdA'.
\ee
   Apparently the integral equation Eq.(\ref{INTEGRAL}) having a kernel
of the Newton potential is the same as that of the equivalent simple
layer for the Dirichlet problem of the Laplace equation.
   Let us relate it to the diffusion equation in the followings.

\par
   Because the order parameter deviation, $\delta s$, from the saturated
level $\pm s_0$ at each phase is expected to be very small in the region
out of the boundary layer ($|z|>\xi$), the chemical potential can be
approximated by
\be
\mu \simeq \DDF{F}{s} \Big|_{\pm s_0}\delta s = {\chi_0}^{-1}\delta s .
\ee
Therefore, the limiting process in this region must be an ordinary diffusion
process defined by
\be \label{DIFF}
\DF{}{t} \psi = D\NN \psi,~~~(D=L/kT\chi_0)
\ee
where a normalized variable, $\psi=\delta s/s_0$, is introduced.
   By using a kind of the boundary layer method we find a boundary condition,
\be \label{GT}
  2s_0 \mu(0) = \sigma H(\mib{a}),
\ee
that is, the well-known Gibbs-Thomson condition for the curved interface
having the mean curvature $H=-\N\cdot\mib{n}$, where $\mib{n}$
is the normal unit vector of the interface.
   Then, the boundary value for $\psi$ must be given by
\be \label{BC}
\psi_\RS (\mib{a}) = \frac{1}{3}\xi H(\mib{a}) .
\ee
   It should be noted that the boundary S deforms itself via this diffusion
with a normal velocity,
\be \label{VELOCITY}
 v_n(\mib{a})=\frac{D}{2}\triangle\big[\mib{n}\cdot\N\psi\big]_{\mib{a}} ,
\ee
where $\triangle[...]_{\mib{a}}$ denotes
a gap of the enclosed quantity through the boundary layer at $\mib{a}$.
   These equations make a Stefan problem with a time-dependent, autonomous
boundary condition.
\par
   In the late stage of the phase separation process where the
characteristic length, $\lambda$, of the spatial pattern is sufficiently
greater than the interface thickness $\xi$, the propagating velocity 
$v_n(\sim D\N\psi\sim D\xi H/\lambda\sim D\xi/\lambda^2)$ and the diffusion
velocity
$v_{\rm D}(\sim D\N \psi/\psi\sim D/\lambda)$ satisfy the condition
\be
 \frac{v_n}{v_{\rm D}} \sim \frac{\xi}{\lambda} \ll 1. 
\ee
Therefore, a quasi-static approximation,
\be
\NN \psi = 0,
\ee
is applicable in this stage.
   Thus, the above Stefan problem becomes a Dirichlet problem of the Laplace
equation.
   The propagating velocity Eq.(\ref{VELOCITY}) together with the boundary
condition Eq.(\ref{BC}) is now given by the gap of the `electric field'
across the boundary, i.e., the equivalent `simple charge layer', 
which should satisfy the integral equation Eq.(\ref{INTEGRAL}). 
   The additional parameter $\overline{H}$ should be regarded as a
compatibility condition in this framework as follows:
   When $\{\psi(\mib{r})\}$ is the solution for the boundary condition
$\{\psi_{\RS}(\mib{a})\}$, $\{\psi(\mib{r})+c\}$ is the solution for 
another boundary condition $\{\psi_{\RS}(\mib{a})+c\}$ with an arbitrary
additional constant $c$.
   Obviously the family of the boundary conditions $\{\psi_{\RS}(\mib{a})+c\}$
have the same value of the field gap $\triangle[\mib{n}\cdot\N\psi]_{\mib{a}}$
or the simple layer solution.
   If the equivalent simple layer solution for this family exists, it
determines a unique boundary value given by the left hand side
of the integral equation.
   Then the right hand side should have an adjustable constant for
the compatibility when an arbitrary boundary condition
$\{\psi_{\RS}(\mib{a})\}$ is given.
   Note that this is true only if no flux lines escape out of the 
system into infinite points as is mentioned below.
   That corresponds to the conservation condition
Eq.(\ref{CONSERVATION}), because of the Gauss theorem, i.e., the total
escaping flux being given by the total {\it charge},
\be
   \intS v_n(\mib{a}) dA ,
\ee
included in the system.
   Thus the constant $\overline{H}$ must be determined by the
conservation condition again.
   An exceptional example not covered by this rule is a problem
of an isolated system composed of finite closed surface(s),
 where the flux may escape out.
   Of course we know that this is an ordinary case in the usual
electro-static problem.
   For the simplicity suppose a finite sphere of radius $R$ and a constant
boundary value $\psi_{\RS}$ on it.
   The simple layer in this case is given by
$\triangle[\mib{n}\cdot\N\psi]=\psi_{\RS}/R$, because $\psi(r)=R\psi_{\RS}/r$
for $r\ge R$ and $\psi(r)=\psi_{\RS}$ for $r<R$. 
   Evidently, another boundary condition, $\psi_{\RS}+c$, has a different
value of the simple layer, $(\psi_{\RS}+c)/R$.
   Thus, for the problem of isolated closed surface(s), the compatibility
constant is not necessary. 
   In contrast to it, for the extraordinary case of infinitely extended
systems such as a system of spheres scattered homogeneously in the whole
$\mib{R}^3$ space or a bicontinuous system spreading over the whole
$\mib{R}^3$ space, etc., we need the compatibility condition.
   In these systems the constant boundary values induce a unique constant
potential in {\em both} sides of the interface(s).


\section{Derivation of the Explicit Solution}
\subsection{A planar boundary}
   First let us consider a Dirichlet problem of the Laplace equation,
\be
\NN \psi = 0 ,
\ee 
when the values $\{\psi_{\RS}(x,y)\}$ on the $xy$-plane, S, are given.
   Let $\sigind(x',y';z)$ be the surface charge density on the
grounded conductor S induced by a probe charge $-1$ located at
a referred point $(0,0,z)$.
   Then the potential $\psi$ at the referred point $(0,0,z)$ is expressed
in an explicit form as
\be \label{IND}
 \psi(0,0,z)=\int\!\!\int \sigind(x',y';z)\psi_{\RS}(x',y')dx'dy',
\ee
by using the elementary method of the Green function.
   The induced charge density $\sigind$ on the plane boundary can be found
in every textbook on electro-statics as the simplest example of the mirror
image method.
   By using the actual expression for it together with the Taylor expansion,
\be
\psi_{\RS}(x',y')= \exp(\mib{a}'\cdot\NS)\psi_{\RS}(0,0),
\ee
where $\mib{a}'=(x',y')$ and $\NS=(\D/\D x,\D/\D y)$, Eq.(\ref{IND}) is
rewritten as
\ba \label{BESSELINT}
\psi(0,0,z) &=& \int\!\!\int \frac{z}{2\pi(x'^2+y'^2+z^2)^{3/2}}
\exp(\mib{a}'\cdot\NS)\psi_{\RS}(0,0)
dx'dy'\nn\\
&=&\int_0^{\infty}da'~\frac{za'}{(a'^2+z^2)^{3/2}}J_0(a'|i\NS|)~
\psi_{\RS}(0,0)\\
&=&\exp(-z|i\NS|)~\psi_{\RS}(0,0) \nn,
\ea
where $J_n(r)$ is the usual Bessel function of $n$-th order.
  Then, by differentiating it with respect to $z$, we find
\be
  \mib{n}\cdot\N \psi \big|_{z=+0} = -|i\NS|\psi_{\RS}(\mib{a}).
\ee
   Note that this result itself means merely a transform of the boundary
condition from a Dirichlet type into a Neumann type.
   Combining with the same, symmetric result for the opposite side, 
$z\rightarrow -0$, we obtain the formula for the equivalent simple layer,
\be \label{EQSIGMA}
   \sigma_{\rm eq}(\mib{a})=-\triangle\big[ \mib{n}\cdot\N \psi \big]_{\mib{a}} 
=2|i\NS|~[\psi_{\RS}(\mib{a})-\overline{\psi_{\RS}}] ,
\ee
by means of which we can write the potential in another explicit form,
\be
\psi(\mib{r})=\intS G_0(\mib{r}-\mib{a}')\sigma_{\rm eq}(\mib{a}')dA'.
\ee
   The compatibility constant mentioned in the last part in \S 2 is
incorporated here, although it would disappear after the differentiation
in the present case.
   It should be noted that, from the view-point of the Green function method
this term is required, in principle, for the condition $\psi\rightarrow 0$
when $|z|\rightarrow\infty$ on applying the Green theorem.

   The fictitious operator $|i\NS|$ was introduced by Ohta and Nozaki\cite{ON}
in a perturbational theory for a planar interface.
   They defined it by a Gaussian integral.
   However, in the context of the present derivation it should be 
interpreted as the following limit,
\be
   |i\NS| =\lim_{\lambda\rightarrow\infty}\frac{1}{\lambda}V(-\lambda^2\NNS),
\ee
where $\NNS~(=\D^2/\D x^2+\D^2/\D y^2)$ is the surface Laplacian and the
function $V(Q^2)$ is defined by
\ba \label{V}
 V(Q^2) &=&-\sum_{n=0}^{\infty}\frac{1}{(2n-1)[(2n)!!]^2}(-Q^2)^n\nn\\
        &=& J_0(Q)-QJ_1(Q)+Q\int_0^{Q}J_0(Q')dQ' .
\ea
\par
   The asymptotic behavior for large $|Q|$ is evaluated as
\be \label{ASYMPTO}
V(Q^2)\sim |Q| ,
\ee
which results in the well-known $|q|^3$ dispersion 
relation\cite{JZ}-\cite{SO} for the relaxation mode on a planar interface
obeying the decay law $\sim\exp(-D\xi q^3 t/3)$.
   The formula Eq.(\ref{V}) was derived in the previous paper by introducing
a cut-off length or the upper limit $\lambda$ for the integration in 
Eq.(\ref{BESSELINT}). 
   This time the cut-off is temporarily required to integrate the
series expansion for the Bessel function $J_0(a'|i\NS|)$, or for
$\exp(\mib{a}'\cdot\NS)$, term by term.
   It will be regarded as a physical screening length and has an essential
role in the final formula for bicontinuous systems in the followings.
\par
   The above formulation is applicable to the two dimensional system also,
where the function $V(Q^2)$ should be replaced by
\ba
V(Q^2)&=& -\frac{2}{\pi}\sum_{n=0}^{\infty}\frac{1}{(2n-1)(2n)!}(-Q^2)^n\nn\\
      &=& \frac{2}{\pi}\big[\cos Q+Q\int_0^Q\frac{\sin Q'}{Q'}dQ'\big]~,~(d=2).
\ea
   The asymptotic behavior is the same as Eq.(\ref{ASYMPTO}).
   Here it should be noted that the operator $|i\D/\D x|$ in this case is
neither $\D/\D x$ nor $|\D f/\D x|$ if operated on $f(x)$.
\subsection{General curved surface boundaries}
   Now let us discuss the general case of a parametric, curved surface S.
   Let $(u_1,u_2,u_3)$ be a curvilinear orthogonal coordinate system, where
the surface S is defined by $u_3=0$.
   The Laplacian in this system is given by
\be \label{LAPLACE}
\NN =\frac{1}{\gamma}\sum_{i=1}^{3}\DF{}{u_i}\frac{\gamma}{{g_i}^2}\DF{}{u_i},
\ee
where
\be
g_i=\big|\DF{\mib{r}}{u_i}\big| = 1/|\N u_i|,
\ee
is the linear metric ($i$-th element of the diagonalized metric tensor) and
$\gamma$ is the Jacobian defined by $\gamma=g_1g_2g_3$.
   We need the infinitesimal vicinity of S only in the followings.
   Then it is always possible without loosing the generality to assume
$g_3(u_1,u_2,u_3)=1$ everywhere by introducing a family of
parallels\cite{WEATH} of S for the simplicity.
   In addition, let $(u_1, u_2)$ be the orthogonal coordinate associated
with the directions of principal curvature on S.
   Suppose that a probe charge $-1$ is located at a referred point
$(0,0,z)$ in this curvilinear coordinate and the surface S is grounded.
   The Green function $G(u_1, u_2, u_3; z)$ of the Laplace equation for
this boundary condition obeys the Poisson equation,
\be \label{POISSON}
 \sum_{i=1}^{3}\DF{}{u_i}\epsilon_i\DF{}{u_i}~G = 
           \delta(u_1)\delta(u_2)\delta(u_3-z),
\ee
with the constraint,
\be
 G(u_1,u_2,0;z) = 0 .
\ee
Here new parameters $\epsilon_1=g_2/g_1, \epsilon_2=g_1/g_2$ and 
$\epsilon_3=g_1g_2$ are introduced for convenience.
Note that the denominator $\gamma$ in Eq.(\ref{LAPLACE}) canceled out with
the Jacobian, which had appeared in the right hand side of
Eq.(\ref{POISSON}).
   This can be regarded as an electro-static problem with a planar boundary
condition in a {\it Euclidean~} space $(u_1, u_2, u_3)$ having an anisotropic,
heterogeneous dielectric tensor, {\it if} the whole parametric space
$(u_1, u_2, u_3)$ corresponds to the original physical space, and vice versa.
   For the time being let us assume it on condition that we need only the
limit $z\rightarrow 0$, and let it be discussed later.

\par
   Therefore, the elementary mirror image method is applicable to this case,
taking account of the refraction of the flux lines,
\be
 \mib{D}=\Big(-\epsilon_1 \DF{G}{u_1},-\epsilon_2 \DF{G}{u_2},
-\epsilon_3 \DF{G}{u_3}\Big) .
\ee
   For the present purpose to find $\mib{n}\cdot\N\psi$ on S we need only
the flux lines that run along the infinitesimal vicinity of S in order to
calculate the induced charge $\sigind$, because the fields
produced by the probe charge at $z$ and by its image at $-z$ can be
calculated separately.

\par
   The method of a reference frame used in the previous paper to calculate
the refracted flux assuming a minimal surface can be extended to more
general surfaces as follows: 
   If we take a normalization, $\epsilon_i(0,0,z)=1$, for the
simplicity, the flux that arrives in the area element $du_1du_2$ on the
{\it plane} S is related to the solid angle $d\Omega$ when the flux started
from the source point $(0,0,z)$, that is, $D_3 du_1du_2 = d\Omega/4\pi$.
   Let $d\tu_1d\tu_2$ be the area element that $d\Omega$ would cut from
the {\it plane} S when it were extended straight without refraction.
   Then the solid angle $d\Omega$ is given by
\be
 d\Omega =\frac {z}{({\tu_1\!}^2+{\tu_2\!}^2+z^2)^{3/2}}d\tu_1d\tu_2 .
\ee
   Thus, the required surface charge induced in the area element $dA$ on
 the original curved surface S is finally given by
\be
\sigind(u_1,u_2;z) dA=-\DF{G}{u_3}\Big|_{u_3=0} dA
= \frac {z}{2\pi({\tu_1\!}^2+{\tu_2\!}^2+z^2)^{3/2}}d\tu_1d\tu_2 ,
\ee
where the relation $dA=\epsilon_3 du_1du_2$ is used
and a factor 2 caused by the image charge is introduced.
   Note that the actual expression for the refraction law of the flux, or
that for the mapping $(u_1,u_2)\rightarrow (\tu_1,\tu_2)$ are not necessary
in the above calculations.

\par
   This is the method of the pseudo-conformal transformation introduced
in the previous paper.
   Thus the assumption of a minimal surface used there has none of special
meaning for us no longer.

\subsection{The bicontinuous phase}
   The remaining procedure is almost the same as that for the planar
boundary except for the following three facts:
   First, it should be noted that the operator we will face in the 
expected formula related to Eq.(\ref{BESSELINT}) after the Taylor
expansion must be
\be \label{LAPTILDE}
     \DDF{}{\tu_1\!}+\DDF{}{\tu_2\!}.
\ee
   However, in the limit $z\rightarrow 0$, this can be replaced by the
desired surface Laplacian defined by
\be
\NNS = \frac{1}{g_1g_2}\Big(
\DF{}{u_1}\frac{g_2}{g_1}\DF{}{u_1}+\DF{}{u_2}\frac{g_1}{g_2}\DF{}{u_2}
\Big) ,
\ee
as is shown in Appendix A.
   Note that this fact does not mean a conformal mapping as is mentioned there.

\par
   Second, there still remains the question whether the whole physical space
is covered by a parametric space $(u_1, u_2, u_3)$ or not.
   Apparently the answer is no in general except for simple curved surfaces
having the same topology as that of a plane.
   In a complicated sponge structure the region described by a given set of
parameters $(u_1, u_2, u_3)$ must be limited in the local {\it cave}
around the referred point.
   In order to avoid this difficulty, let us introduce an upper-limit
$\lambda$ for the integration in Eq.(\ref{BESSELINT}).
   This cut-off length was defined intuitively as an {\it electro-static}
screening length, i.e., the effective diameter of local caves
of the sponge-shaped {\it conductor} in the previous paper.
   In fact we have such a unique (but time-dependent), well-defined
characteristic length in the bicontinuous phase that grows up starting from
a quenched, homogeneous mixtures.
   It shou1d be noted that we have another difficulty that at least one of
the parameters $(u_1,u_2)$ may become multi-valued, for example,
on a part of the surface having a rotational symmetry like a catenoid.
   It seems that we have no problem in applying the image method to this
case {\it if} we do not introduce any cut-off, as is exemplified using
a two-dimensional solvable problem of a circle boundary in Appendix B.
   The cut-off procedure, however, causes literally cut-off of the multi-valued
part in this case.
   Fortunately, in addition to the fact that we need only the very vicinity
of S, the weight of the contribution of the boundary values at the points 
away from the referred point decreases as $\sigind\sim 1/\tilde{r}^{3}$,
where $\tilde{r}=({\tu_1\!}^2+{\tu_2\!}^2)^{1/2}$ is essentially
the distance along S as is suggested by the equality between the operator
Eq.(\ref{LAPTILDE}) and $\NNS$.
   Thus, the cut-off procedure may be introduced without ruining the
approximation.

\par
   The last point is the bicontinuous nature of the interface S.
   This is related to the equivalence of the boundary condition between
the inner and the outer problems of S.
   For example, let us consider a non-bicontinuous system composed of
several closed surfaces, $\RS_1,\RS_2,...$~.
   Evidently, each inner problem associated with each $\RS_i$ has a single
boundary $\RS_i$ itself.
   On the other hand the sole outer problem has a united boundary
 $\RS_1+\RS_2+...$~.
   Thus the boundary conditions are not equivalent to each other.
   In contrast to it, we have just two regions separated by a single sheet
of the interface S in the bicontinuous phase that is extended infinitely or
is connected periodically in all three directions, and the difference between
the `inner' and `outer' problems has no sense there.
   Both problems have exactly the same boundary in this case.
   Then the induced charge densities for both problems must coincide with
each other when $z\rightarrow\pm 0$ as is seen in the above derivation,
even if the both bulk regions are asymmetric at an average.
   Therefore, the values of the potential gradient
$\mib{n}\cdot\N\psi(\mib{a})$ constructed with the common boundary values
$\{\psi_{\RS}(\mib{a})\}$ on S become equivalent one another.

\par
   Thanks to these facts we can construct the equivalent simple layer
by using Eq.(\ref{EQSIGMA}).
   Thus the explicit expression for the curvature flow equation,
\be \label{CFE}
   v_n(\mib{a})=-\frac{D\xi}{3\lambda(t)}V(-\lambda(t)^2\NNS)~[H(\mib{a})-
\overline{H}],
\ee
is obtained again for general bicontinuous phases.
   The parameter $\lambda(t)$ is the time-dependent characteristic length,
which is to be determined self-consistently by this equation.
   On the assumption of scaling, i.e., the time-dependent similarity law,
the well-known time-dependence $\lambda(t)\sim t^{1/3}$ may be found
by using a kind of dimension analysis on this equation.
   Note that the compatibility constant $\overline{H}$ in the right hand side
is given by the simple surface average in this approximation, i.e.,
\be
\overline{H}=\intS H(\mib{a})dA\Big/ \intS dA .
\ee

   For small $Q$, the function $V(Q^2)$ is expanded as\cite{TPRO}
\be
V(Q^2)\simeq\left\{ \begin{array}{ll}
                   1+\frac{Q^2}{4}+... & (d=3), \\
                   \\
                   \frac{2}{\pi}\big(1+\frac{Q^2}{2}+...) & (d=2) .
                    \end{array}
            \right.
\ee
\begin{figure}
   \epsfxsize=2.5in
   \centerline{\epsfbox{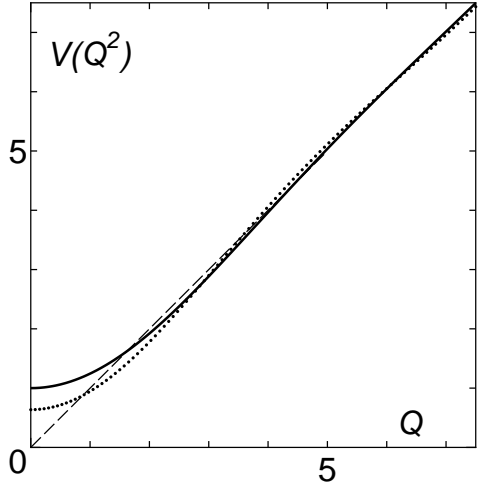}}
   \caption{The function $V(Q^2)$ for $d=3$ (solid line) and
            $d=2$ (dotted line).}
   \label{fig:1}
\end{figure}
   The first term expresses the mean-field evaporation-condensation process 
$(\sim(H(\mib{a})-\overline{H}))$, due to the imbalance of the mean curvature
around its average value $\overline{H}$.
   The second term may be interpreted as a kind of surface diffusion 
$(\sim - \NNS H(\mib{a}))$ due to the local nonuniformity of the curvature.
   On the contrary the asymptotic value of $V(Q^2)$ for $Q\rightarrow \pm
\infty$ is given by Eq.(\ref{ASYMPTO}), i.e., $V(Q^2)\simeq |Q|$, which
corresponds to the $|q|^3$ relaxation mode $(\sim -|\NS|H(\mib{a}))$.
   Then the short wave-length local fluctuations ($|q|\lambda(t)\gg1$)
on the interface must decay rapidly and we can expect that the interface
is always smooth within a spatial scope of $\lambda(t)$ almost everywhere.
\par
   The above behaviors of $V(Q^2)$ are shown in Fig.1.
   The asymptotic estimation ($\sim|Q|$) is practically applicable for 
$Q>2$.
   However, the important modes for the interface evolution are included
in the small wave-number region, $|q\lambda(t)|<1$, where $V(Q^2)$
deviates significantly from its asymptote.
\section{Discussions}
   The Onsager principle of minimum dissipation seems to be really useful
in such a practical problem to reduce the interface evolution equation from
the bulk equation, at least if the effect of thermal fluctuations can be
neglected.
   In the unstable phase separation process the main role of thermal
fluctuations is creating nuclei or microscopic droplets at the very early
stage.
   This effect may be incorporated as the initial condition in the picture
of the interface dynamics.
   Another example is the NCOP system.
   The inverse of the Onsager coefficients in this case is merely
$L^{-1}\delta(\mib{r}-\mib{r}')$.
   Then the dissipation function is given by
\be
   \Phi=\frac{1}{2L}\int \big(\DF{s(\mib{r})}{t}\big)^2 d\mib{r}
       =\frac{\chi_0 \sigma}{2L\xi^2}\intS v_n(\mib{a})^2 dA ,
\ee
where the definition of the surface tension Eq.(\ref{TENSION}) is used.
   On the other hand, $\dot{S}$ becomes
\be
\dot{S}=-\frac{1}{kT}\intS\Big\{\sigma H(\mib{a})-2s_0 h\Big\}v_n(\mib{a})dA,
\ee
where $h$ is the uniform external field that should be included in the free
energy in the following form,
\[
  -\int hs(\mib{r}) d\mib{r},
\]
for a NCOP system.
   Then using the variational method we find
\be
   v_n(\mib{a})=-\frac{L\xi^2}{kT\chi_0}\big[H(\mib{a})-\frac{2s_0h}{\sigma}
\big] .
\ee
   That is the well-known Allen-Cahn equation.\cite{AC}\ 
   In addition, putting $H=(d-1)/R$ for the droplet system, we obtain a radius
evolution equation,
\be
   \dot{R}=\frac{(d-1)L\xi^2}{kT\chi_0}\Big(\frac{1}{R_c}-\frac{1}{R}\Big),
\ee
where $R_c=(d-1)\sigma/2s_0h$ is the radius of the critical droplet.
   Thus we can get these important formulae straightforwardly and correctly
by using the Onsager principle.

\par
   The one-parameter curvature flow equation for general COP systems is
rederived without assuming a minimal surface.
   The bicontinuous nature is shown to be indispensable in the present
derivation.
   Of course the result is applicable to a planar interface if we adopt
an infinite cut-off length, $\lambda\rightarrow \infty$. 
   Especially it should be noted that it gives an exact explicit
solution of the Dirichlet problem for an arbitrary curved surface having
the same topology as that of a plane.
   Such formulae have not been found in the textbook on the electro-statics
or on the potential problem.
\par
   On the contrary, this result cannot be applied to a system of
scattered spheres because it has not the bicontinuous nature,
although it has been one of the simplest examples to study.
   Of course a more accurate argument on this system\cite{PHYSICA} is possible
by applying the standard electro-statics directly to
Eq.(\ref{INTEGRAL}) as was performed by Kawasaki and Ohta.\cite{KOP}\ 
   Suppose each sphere of radius $R_i$ and charged by $Q_i$ is
sufficiently separated from each other.
   Then Eq.(\ref{INTEGRAL}) for the potential on the surface
of the sphere $i$ becomes
\be
\frac{Q_i}{4\pi R_i}+\sum_{j\ne i}\frac{Q_j}{4\pi R_{ij}} 
=\frac{D\xi}{6}\big[\overline{H}-\frac{2}{R_i}\big],
\ee
where $R_{ij}$ is the distance between a couple of spheres $i$ and $j$
and is assumed as $R_{ij}\gg R_i, R_j$.
   Then the surface charge density given by $\dot{R}_i=Q_i/4\pi {R_i}^2$
satisfies
\be
\dot{R}_i+\sum_{j\ne i}\frac{{R_j}^2}{R_i R_{ij}}\dot{R}_j
=\frac{D\xi}{3}\frac{1}{R_i}\big[\frac{1}{R_c(t)}-\frac{1}{R_i}\big].
\ee
   The second term in the left-hand side shows plainly the long-range
interaction of the interface through the diffusion process.
   When this term is neglected for a sufficiently thin system, this
 becomes the Lifshitz-Slyozov equation,\cite{LS}\ where the critical
radius $R_c(t)$ in the right-hand side is to be determined by
\be
  R_c(t)=\langle R\rangle =\sum_{i} R_i/\sum_{i} 1,
\ee
according to the conservation condition $\sum_{i}4\pi {R_i}^2\dot{R}_i=0$
in this system.

\par
   Lastly let us discuss the equation of the level function associated with 
the present curvature flow equation.
   This method was used first by Ohta {\it et al}\cite{OJK} in a special
problem of the ordering process in a NCOP system with $h=0$ and has been
developed by mathematicians\cite{LEVEL} independently:
   Let $\{u(\mib{r},t)\}$ be the fictitious scalar field, the interface
being defined by $u(\mib{r},t)=0$.
   Then the interface normal velocity is represented as
$v_n=|\N u|^{-1}(\D u/\D t)_{u=0}$.
   Suppose we have an explicit curvature flow equation,
\be
   v_n(\mib{a})={\cal F}(\{H(\mib{a}')\}) ,
\ee
where ${\cal F}$ may be a functional.
   Here the mean curvature is given by
\be
   H = -\N \cdot\mib{n} =-\big[ |\N u|^{-1}\NS\cdot\N u\big]_{u=0} ,
\ee
where $\NS=\N-\mib{n}(\mib{n}\cdot\N)$.
   If we extend these equations to the bulk ($u\ne 0$), that makes a closure of
$\{u(\mib{r},t)\}$.

\par
   Thus the level function equation for our case is written as
\be \label{U}
\DF{u}{t} =\frac{D\xi}{3\lambda(t)}~|\N u|~V(-\lambda(t)^2\NNS)
\big[ |\N u|^{-1}\NS\cdot\N u+\overline{H}\big] .
\ee
   For a small value of $u$ we have a formula,\cite{WEATH}
\be \label{HU}
H(u)-H(0) \simeq u~\big[|\N u|^{-1}(H^2-2K)+\NNS|\N u|^{-1}\big]_{u=0} ,
\ee
where $K$ is the Gauss curvature defined by $K=1/R_1 R_2$ using the radii of
the principal curvature, $R_1$ and $R_2$.
   Then $\overline{H}$ in the right hand side of Eq.(\ref{U}) may be replaced
by
\be
\overline{H}=\overline{H(0)}+u~\overline{\big[|\N u|^{-1}(H^2-2K)\big]}_{u=0} .
\ee
   The second term in the right hand side of Eq.(\ref{HU}) canceled out after
the surface integration.
   Note that the quantity $H^2-2K~(={R_1}^{-2}+{R_2}^{-2})$ is positive definite.
   Especially for a homogeneous bicontinuous phase of a symmetric mixture,
this can be estimated as
\be
   H^2-2K \simeq -2K \simeq 2\lambda(t)^2 ,
\ee
or alternatively we may use it as a definition of the parameter $\lambda(t)$.
\par
   Thus the level function equation for a COP system has very tough form and
has several difficulties\cite{YO} to be solved.
   None of satisfactory analyses on it has been found so far.\cite{ON,TPRO,MAZ}

\section*{Acknowledgements}
The author would like to thank the members of the statistical physics group of
Nara Women's University, who invited him to have a lecture on the phase 
ordering process in the autumn of 1999 and stimulated him into the present
work in preparing a lecture note.

\appendix
\section{on the pseudo-conformality}
   Let $f(u_1,u_2)$ be an arbitrary continuous function on S and suppose
a harmonic function satisfying the Laplace equation
\be \label{A1}
\Big( \NNS+\frac{1}{\epsilon_3}\DF{}{u_3}\epsilon_3\DF{}{u_3}\Big)
\psi(u_1,u_2,u_3)=0 ,
\ee
with the boundary condition
\be \label{A2}
\psi(u_1,u_2,0)=f(u_1,u_2) .
\ee
   Let us generalize the reference frame on S defined in \S 3 to the bulk,
i.e., to the region $u_3\ne 0$ in the same manner.
   Then by using the Green function method a formal solution for the 
harmonic function $\psi$ is given by 
\be \label{A3}
\psi(u_1,u_2,u_3)=-\intS \DF{\tilde{G}}{u_3'}\Big|_{u_3'=0}
f(\mib{a}')d\tilde{A}' ,
\ee
where
\be \label{A4}
\tilde{G}(\tu_1',\tu_2',u_3';\tu_1,\tu_2,u_3)=\tilde{G}^{+}-\tilde{G}^{-},
\ee
and
\be \label{A5}
\tilde{G}^{\pm}=\frac{1}{4\pi[(\tu_1'-\tu_1)^2+(\tu_2'-\tu_2)^2
+(u_3'\mp u_3)^2]^{1/2}} ,
\ee
which satisfy a Poisson equation
\be \label{A6}
\Big(\DDF{}{\tu_1\!}+\DDF{}{\tu_2\!}+\DDF{}{u_3\!}\Big)\tilde{G}^{\pm}=
-\delta(\tu_1'-\tu_1)\delta(\tu_2'-\tu_2)\delta(u_3'\mp u_3) .
\ee
   Note that the reference frame is defined on taking the origin at the point
just under the probe charge in \S 3.
   However, once the new {\it Euclidean} coordinate system is defined, one may
shift the origin to any point.
   Thus, the reference coordinate $(\tu_1,\tu_2, u_3)$ is also used for the
point of the probe charge in the above equations.

\par
   Now combining Eqs.(\ref{A1}),(\ref{A3}) and (\ref{A6}) it is
straightforward to find
\be \label{A7}
   \Big(\DDF{}{\tu_1\!}+\DDF{}{\tu_2\!}\Big) f = \NNS f ,
\ee
in the limit, $u_3\rightarrow 0$.
   Here the following relation for small $u_3$,
\be \label{A8}
\frac{1}{\epsilon_3}\DF{}{u_3}\epsilon_3\DF{\tilde{G}^{\pm}}{u_3}
=H\DF{\tilde{G}^{\pm}}{u_3}+\DDF{\tilde{G}^{\pm}}{u_3\!}
\simeq\DDF{\tilde{G}^{\pm}}{u_3\!} ,
\ee
is used. 
   Of course this approximation is exact for a minimal surface ($H=0$).
   In general one may use the fact that $\tilde{\sigma}
=-(\D\tilde{G}^{\pm}/\D u_3')_{u_3'=0}$, which has an almost $\delta$-function
shape, satisfies
\be
   \DF{\tilde{\sigma}}{u_3}\sim\frac{1}{u_3}\tilde{\sigma},~~
\DDF{\tilde{\sigma}}{u_3\!} \sim\frac{1}{{u_3}^2}\tilde{\sigma},
\ee
and so on, when $u_3$ is sufficiently small.

\par
   Using the same argument repeatedly on the functions $\NNS f, {\NS\!}^4 f$
..., we obtain
\be
   \Big(\DDF{}{\tu_1\!}+\DDF{}{\tu_2\!}\Big)^{n}f={\NS\!}^{2n}f,
\ee
for an arbitrary integer $n$.
   This is the result desired in \S 3. 
\par
   It should be noted that this does not mean a conformal transformation on S,
because the definition of the reference coordinate itself depends on the
probed point.
   Further the reference coordinate $(\tu_1,\tu_2)$ itself is not
an orthogonal system on S in general.
   Then, strictly speaking, the operator
$\D^2/\D{\tu_1\!}^2+\D^2/\D{\tu_2\!}^2$ is not an invariant Laplacian
but merely a Taylor expansion operator around the referred point.

\section{a solvable example in $d=2$}
   It is instructive to discuss a solvable problem in the present scheme,
though it is stupid practically.
   Let us consider a circle of radius $R$ and assume a probe charge is located
at a distance $z$ from it.
   A new curvilinear coordinate $(u_1,u_2)$ may be defined by
\be \label{B1}
  u_1=(R+z)\varphi,~~u_2=r-R ,
\ee
where $(r,\varphi)$ is the usual polar coordinate.
   The two dimensional Laplacian is written in this coordinate as
\be \label{B2}
  \NN =\frac{1}{g_1}\Big(\DF{}{u_1}\frac{1}{g_1}\DF{}{u_1}+
       \DF{}{u_2}g_1\DF{}{u_2}\Big) ,
\ee
where 
\be \label{B3}
   g_1=|\N u_1|^{-1}=\frac{R+u_2}{R+z}~,~~g_2=|\N u_2|^{-1} =1 .
\ee
\par
   Let us derive the refraction law of the flux $\mib{D}$ in the
{\it Euclidean} space $(u_1,u_2)$ due to the anisotropic dielectric tensor
$(\epsilon_1,\epsilon_2)=({g_1}^{-1},g_1)$, which depends on $u_2$ only.
   Let $\alpha$ be the angle between the flux line and the $u_1$-axis.
   Then we find that
\be \label{B4}
   \epsilon_1\tan\alpha = {\rm constant} ,
\ee
because of the continuity relations across a boundary line parallel to the
$u_1$-axis,
\be \label{B5}
   D\sin\alpha =D'\sin\alpha'~,~~ \frac{D\cos\alpha}{\epsilon_1}
=\frac{D'\cos\alpha'}{\epsilon_1'},
\ee
for the perpendicular and the parallel components, respectively.
   By using Eq.(\ref{B4}) the flux equation is given by
\be \label{B6}
   \frac{du_2}{du_1}=-\frac{\tan\alpha_0}{\epsilon_1}
=-\frac{R+u_2}{R+z}\tan\alpha_0 ,
\ee
or
\be \label{B7}
   \log(R+u_2)=\log(R+z)-\frac{\tan\alpha_0}{R+z}u_1 ,
\ee
where $\alpha_0$ is the initial angle, which defines the reference
coordinate by
\be \label{B8}
 \tu_1 =z/\tan\alpha_0 .
\ee
   On putting $u_2=0$ in Eq.(\ref{B7}) we find the mapping, $u_1\rightarrow
\tu_1$, i.e.,
\be \label{B9}
  \tu_1=\frac{z}{(R+z)\log(1+z/R)}u_1=\frac{z}{\log(1+z/R)}\varphi .
\ee
   Of course this is a multi-valued mapping, $(-\pi,\pi)\rightarrow
(-\infty,\infty)$.
   Then the induced charge density is given by
\be \label{B10}
\sigind d\varphi =\frac{1}{\pi}\frac{z}{{\tu_1}^2+z^2}d\tu_1
=\frac{x}{\pi}\sum_{n=-\infty}^{\infty}\frac{1}{(\varphi+2n\pi)^2+x^2}
d\varphi ,
\ee
where a new parameter $x=\log(R+z)$ is used.
   By using some formulae this yields
\ba \label{B11}
  \sigind d\varphi &=& \frac{1}{2\pi}\frac{\sinh x}{\cosh x - \cos \varphi}
d\varphi \nn\\
&=&\frac{1}{2\pi}\frac{(2R+z)z}{R^2+(R+z)^2-2R(R+z)\cos\varphi}d\varphi .
\ea
   That is the well-known result obtained directly with use of a mirror image
method.
\par
   Thus the multi-valued mapping has no problem in the present reference
frame method.
   However, the cut-off procedure causes literally the cut-off of the tail
part of the induced charge density.
   Here it is wrong to conclude that this effect would become infinitesimal
in the limit $z\rightarrow 0$ because the cut-off charge is of
order of $z/\lambda(t)$.
   Note that the final formula for the equivalent simple layer is obtained 
after differentiating with respect to $z$.
   The weight for the contribution from the remote points becomes
$\sim d\tilde{\mib{u}}/\tu^{d}$, lacking the factor $z$.
That estimates $\sim 1/\lambda(t)$ for the cut-off effect.
   Thus the cut-off loss must be recovered by some trick, such as
the mean field term $\overline{H}/\lambda(t)$ in Eq.(\ref{CFE}).

\end{document}